\documentclass[conference,a4paper,10pt]{IEEEtran}
\usepackage{amsmath}
\usepackage{bm}

\usepackage{graphicx}

\ifCLASSINFOpdf
\else
\fi
\usepackage{color}	
\begin{document}
%

\title{Pixel Relationships-based Regularizer for Retinal Vessel Image Segmentation}


\author{\IEEEauthorblockN{Lukman Hakim$^{1}$, Takio Kurita$^{2*}$}
\IEEEauthorblockA{1 Department of Information Engineering, Graduate School of Engineering, Hiroshima University, Higashihiroshima, Japan}
\IEEEauthorblockA{2 Graduate School of Advanced Science and Engineering, Hiroshima University, Higashihiroshima, Japan}
\IEEEauthorblockA{$^{*}$E-mail: tkurita@hiroshima-u.ac.jp}
}


%


\maketitle

\begin{abstract}
The task of image segmentation is to classify each pixel in the image based on the appropriate label. Various deep learning approaches have been proposed for image segmentation that offers high accuracy and deep architecture. However, the deep learning technique uses a pixel-wise loss function for the training process. Using pixel-wise loss neglected the pixel neighbor relationships in the network learning process. The neighboring relationship of the pixels is essential information in the image. Utilizing neighboring pixel information provides an advantage over using only pixel-to-pixel information. This study presents regularizers to give the pixel neighbor relationship information to the learning process. The regularizers are constructed by the graph theory approach and topology approach: By graph theory approach, graph Laplacian is used to utilize the smoothness of segmented images based on output images and ground-truth images. By topology approach, Euler characteristic is used to identify and minimize the number of isolated objects on segmented images.  Experiments show that our scheme successfully captures pixel neighbor relations and improves the performance of the convolutional neural network better than the baseline without a regularization term.  
\end{abstract}


\section{Introduction}
Image segmentation is an important task in the field of computer vision. The task of image segmentation is to classify each pixel in the image based on the appropriate label. Image segmentation applications have been widely used, such as medical imagery (for example, Retinal Vessel Segmentation\cite{Fu2016}, Tumor Segmentation\cite{Havaei2017}, Breast Cancer Detection\cite{Zeebaree2019}), Autonomous Vehicles (Example: Vehicle Detection\cite{Chougula2020}, Pedestrian Detection\cite{Ullah2018}), Counting Objects\cite{Cholakkal2019}, and Farm Industry\cite{Wang2017}. Several Image segmentation techniques have been developed, such as thresholding, k-means clustering, and graph cut methods. 

Recently, many deep learning techniques have been developed for image segmentation that offers high accuracy and deep architecture\cite{Ronneberger2015}\cite{Badrinarayanan2017}. However, the deep learning technique uses a pixel-wise loss function for the training process. Using pixel-wise loss neglected the pixel neighbor relationships in the network learning process. The neighboring relationship of the pixels is essential information in the image. Utilizing neighboring pixel information provides an advantage over using only pixel-to-pixel information. 

\begin{figure}[!t]
    \centering
    \includegraphics[width=8cm]{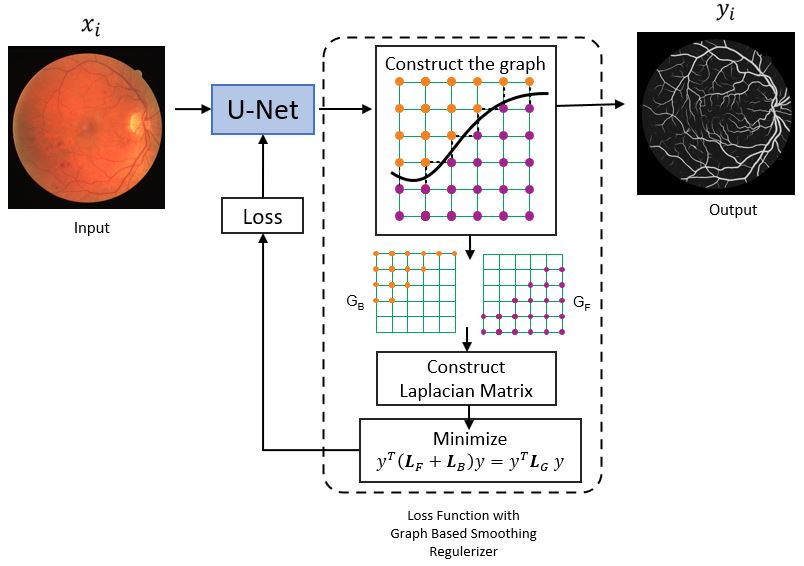}
    \caption{Illustration of graph based smoothing regularizer (GBS) architecture.}
    \label{fig-gbs}
\end{figure}

\begin{figure}[t]
    \centering
    \includegraphics[width=6cm]{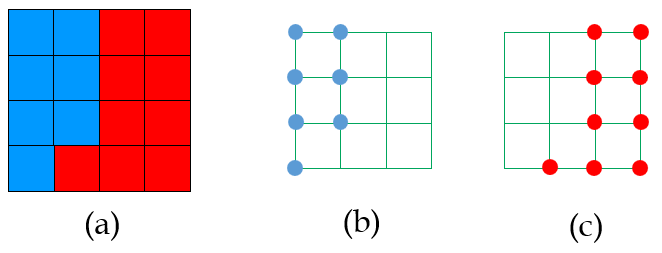}
    \caption{Representation of graph between two regions, a) Ground-truth image, b) graph generated from background region and c) graph generated from foreground region.}
    \label{fig-gbs-ide}
\end{figure}

This study presents regularizers to provide the pixel neighbor relationship information to the learning process. We have proposed three regularizers are constructed by the graph theory approach and topology approach. Firstly, we introduced a Graph-Based Smoothing Regularizer (GBS). The GBS considers the graph laplacian from the foreground and background regions and then combines it with the CNN baseline loss function. The combination of regularizers allows the network to learn the pixel relationship efficiently. The illustration of GBS is shown in Fig. \ref{fig-gbs}. 
\begin{figure}[!t]
    \centering
    \includegraphics[width=8cm]{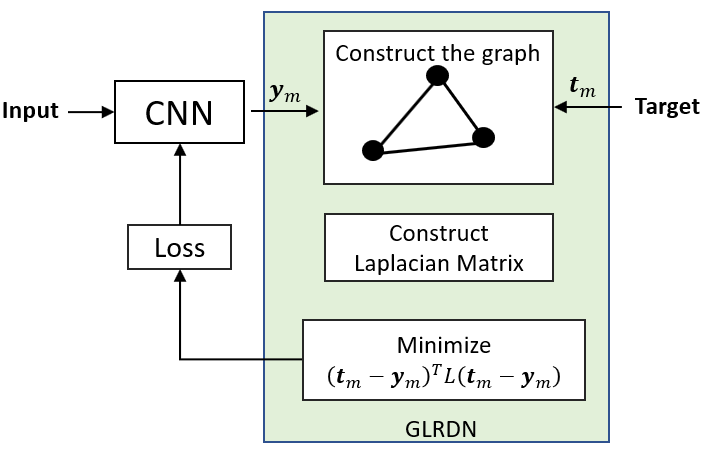}
    \caption{Architecture of Graph Laplacian regularizer based on differences of neighboring pixels.}
    \label{fig-grldn}
\end{figure}

Secondly, we introduced Graph Laplacian Regularization based on the Differences of Neighboring Pixels (GLRDN) by constructing graph laplacian from prediction and ground-truth images. A graph uses pixels as vertices and edges defined by the "differences" of neighboring pixels instead of similarities between pixels. A graph uses pixels as vertices and edges defined by the "differences" of neighboring pixels instead of similarities between pixels. The basic idea is, if pair-wise pixels belonging a similar class, the differences are small. Otherwise, the differences are significant if pair-wise pixels are belonging a different class. The illustration of GLRDN is shown as Fig \ref{fig-grldn}.
\begin{figure}[!t]
    \centering
    \includegraphics[width=8cm]{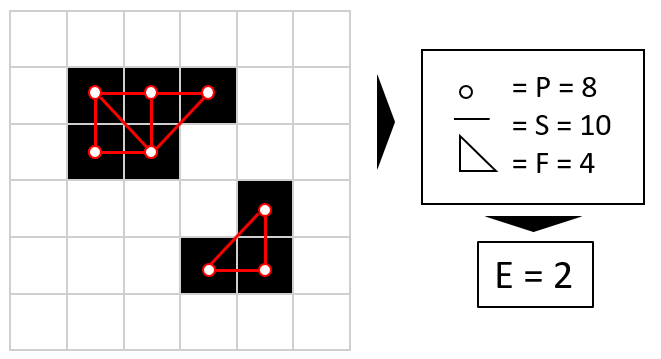}
    \caption{Simple Illustration of Euler Characteristic Calculation.}
    \label{fig-euler}
\end{figure}
Thirdly,  we proposed a regularizer based on Euler Characteristic (EC) in the segmented image. EC is a topological property of a shape in an image that can identify the number of objects in the image. This EC-based regularizer is used to identify the number of objects in the segmentation results. Then this information is added to the objective function so that the network can minimize the number of isolated objects in the segmented image.

To evaluate the effectiveness of the proposed methods, we implemented the regularizers on retinal blood vessels segmentation in fundus images and compared them with the baseline CNN without regularizers.

\section{Methods}
This section explains three proposed regularizers we have done: GBS, GLRDN, and EC Regularizer.

\subsection{Graph Based Smoothing (GBS)}
The proposed graph-based smoothing regularizer is based on the relationship between pixels\cite{hakim2019}. The graph approach is used to obtain adjacency graphs and diagonal matrices from the relationship information between pixels. The graph can be transformed into a matrix by counting the edges between the two adjacent vertices. The two vertices can be said to be adjacent if an edge connects it. Then from this matrix, the Graph Laplacian is defined.

The GBS defines two different graphs for the foreground region and background region. As shown in Fig. \ref{fig-gbs-ide}, let us consider the set of training samples $X=\left(x_m,t_m\right)|m=1,\ldots,M$ where $x_m$ is a $m^{th}$ input image and  $t_m$ is the $m^{th}$  target image, and $M$ is number of training samples.
In deep convolutional neural network, the network is trained to predict the output image ${y}_{m}$ from the $m^{th}$ input image ${x}_{m}$. For each edge of foreground  $\left(j_F,k_F\right) \in E_F$  and background $\left(j_B,k_B\right)\in E_B$ of the graph, the similarity  $\beta_{\left(j_F,k_F\right)}$  and $\beta_{\left(j_B,k_B\right)}$ is defined as 
\begin{equation}
\beta_{(j_F,k_F)} = 1-|t_{j_F}-t_{k_F}|
\end{equation}

\begin{equation}
\beta_{(j_B,k_B)} = 1-|t_{j_B}-t_{k_B}|
\end{equation}

We introduced the regularization term for smoothing $S$ based on foreground region $F$ and background region $B$ as
\begin{equation}
    \sum_{(j_F,k_F) \in G_F}\beta_{j_F,k_F}{(y_{j_F}-y_{k_F})^2}
    = y^T(D_F - A_F)y=y^T L_{F} y
\end{equation}
\begin{equation}
    \sum_{(j_B,k_B) \in G_B}\beta_{j_B,k_B}{(y_{j_B}-y_{k_B})^2}
    = y^T(D_B - A_B)y=y^T L_{B} y
\end{equation}
where, $L_F$ and $L_B$ is Graph Laplacian for foreground region and background region, respectively. The adjacency and diagonal matrices is defined as following 
\begin{equation}
    \binom{A_F=\beta_{(j_F,k_F)},   D_F = \sum_{j_F=1}^N \beta_{j_F, k_F}}{A_B=\beta_{(j_B,k_B)},   D_B = \sum_{j_B=1}^N \beta_{j_B, k_B}}
\end{equation}

Graph based smoothing regularizer $R_{gbs}$ can be written as 
\begin{equation}
    R_{gbs} = y^T(L_{F}+L_{B}) y \\
    = y^{T}L_{G}y
\end{equation}

In these experiments, we use the Binary Cross Entropy (BCE) loss as objective function. The BCE loss is given as below:

\begin{equation}
E_{bce} = \sum_i^M t_i \log(y_i)+(1-t_i)\log(1-y_i)
\end{equation}

where $t$ and $y$ is groundtruth and output, respectively.

The objective function $O$ applied in this study is the summation of the binary cross entropy of each label with the regularization term using graph based smoothing, which is defined as 

\begin{equation}
O_1 = E_{bce} +\lambda R_{gbs}
\end{equation}

Parameter $\lambda$  is used to control the effect of the regularizer.

\begin{equation}
O_2 = E_{bce} +\lambda R_{glrdn}
\end{equation}

\subsection{Graph Laplacian based on Differences of Neighboring Pixels (GLRDN)}

To define the GLRDN, we consider to utilize the differences of the differences of neighboring pixels between the target image and estimated image\cite{hakim2022}. Let assume  target image ${t}_{m}$ and the estimated images ${y}_{m}$, the GLRDN define as

\begin{align}
R_{glrdn}(\bm{t}_m, \bm{y}_m) &= \sum_{(i,j) \in E} \{(t_{mi} - t_{mj}) - (y_{mi} - y_{mj})\}^2 \nonumber \\
&= \sum_{(i,j) \in E} (\Delta t_m^{ij} - \Delta y_m^{ij})^2  \nonumber \\
&= (\Delta \bm{t}_m - \Delta \bm{y}_m)^T (\Delta \bm{t}_m - \Delta \bm{y}_m) \nonumber \\
&= (B \bm{t}_m - B \bm{y}_m)^T (B \bm{t}_m - B \bm{y}_m) \nonumber \\
&= (\bm{t}_m - \bm{y}_m)^T B^T B (\bm{t}_m - \bm{y}_m) \nonumber \\
&= (\bm{t}_m - \bm{y}_m)^T L (\bm{t}_m - \bm{y}_m)
\end{align}
where $B$ is the incident matrix.
It is noticed that $B^TB$ is equal to the graph Laplacian matrix $L$.
Graph Laplacian matrix assumes that the differences of neighbor pixel between target images and estimated images, denoted as $(\bm{t}_m - \bm{y}_m)$ is smooth with respect to concerning corresponding  graph G. In particular, it enforce the value of $(\bm{t}_m - \bm{y}_m)^T L (\bm{t}_m - \bm{y}_m)$ should be small.
In the training process, GLRDN is use with BCE objective function

\subsection{Euler Characteristic-based Regularizer}
The Euler characteristic is a property of an object's shape that is not affected by topological changes such as scale transformations and rotations. EC can define the number of objects in two-dimensional space provided there are no holes in the object. EC can be obtained by calculation of the relations between vertices (P), edges (S), and faces (F ) : 
\begin{equation}
EC=F-S+P	
\label{eq-euler}
\end{equation}

\begin{figure}[!t]
    \centering
    \includegraphics[width=8cm]{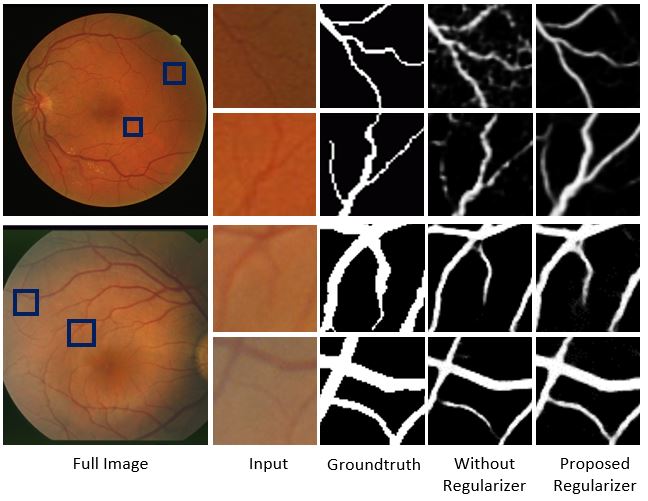}
    \caption{Comparison results of blood vessel segmentation between our proposed regularizer and without proposed regularizer on the DRIVE (top row) and STARE (bottom row) datasets.}
    \label{fig-result}
\end{figure}

For a binary image, the EC can be constructed in the following steps: (a) define pixels as imaginary vertices, (b) calculate the number of edges based on 8x8 neighboring pixels, (c) count the number of vertices, (d) count the number of edges, (e) count the number of faces, (f) calculate the EC using Eq. (\ref{eq-euler}). 

We demonstrate the calculation of Euler Characteristic of simple binary image as shown in Fig. \ref{fig-euler}. In this study, we calculated the EC from the CNN output. Since the EC of an object equals the number of objects in the image, we use this information to determine the number of isolated objects in the CNN output. Then, information about the number of isolated objects is combined with BCE loss as a regularizer in the CNN training process  To minimize the number of isolated objects.

For the EC calculation to be invariant, it is necessary to calculate the vertices, sides, and faces in the opposite direction. Then, calculated the average of the EC in two different directions as follows:

\begin{equation}
R_{ec}=\frac{{EC}_1+{EC}_2}{2}				   
\end{equation}

In this study, EC estimate from the segmentation output of network\cite{hakim2021}.  We combine the objective function with regularization term as:

\begin{equation}
O_3=E_{bce} +\lambda R_{ec}	    
\end{equation}

where $\lambda$ is denoted the parameter to adjust the effect of EC regularizer.

\section{Experiments}
We measure the performance of our proposed method using Sensitivity, Specificity, Accuracy, and Area Under curve on DRIVE, STARE, and CHASEDB1 datasets. To increase the number of datasets, we divide the image into patches of 48 x 48 pixels to get 4,750 patches for each image. In this way, we get a sufficient number of datasets to perform the training step. The first, we compare our proposed method with baseline U-Net architecture with and without regularizer term. For the experiment, we train the network with 100 epochs and initialize the learning rate to 0.001. To achieve convergence, Every time we reach 25 epochs, we reduce the learning rate by 10 times. For the training sample, we set the batch to 32 and used Adam as the optimizer.

\begin{table}[!t]
  \renewcommand{\arraystretch}{1.5}
\caption{Comparison of architectural performance without and with our proposed GBS on the DRIVE datasets.}
\label{tbl-gbs}
\begin{tabular}{l|l|l|l|l|l|}
\cline{2-6}
 &
  \multicolumn{1}{c|}{\textbf{Methods}} &
  \multicolumn{1}{c|}{\textbf{Sn}} &
  \multicolumn{1}{c|}{\textbf{Sp}} &
  \multicolumn{1}{c|}{\textbf{Acc}} &
  \multicolumn{1}{c|}{\textbf{AUC}} \\ \cline{2-6} 
 &
  Baseline &
  0.6707 &
  0.9867 &
  0.9465 &
  0.9652 \\ \cline{2-6} 
 &
  Baseline+GBS &
  0.7064 &
  0.9897 &
  0.9536 &
  0.9794 \\ \cline{2-6} 
\end{tabular}
\end{table}

\begin{table}[!t]
  \renewcommand{\arraystretch}{1.5}
\caption{Comparison of architectural performance without and with our proposed GLRDN on the DRIVE datasets.}
\label{tbl-glrdn}
\begin{tabular}{l|l|l|l|l|l|}
\cline{2-6}
 &
  \multicolumn{1}{c|}{\textbf{Methods}} &
  \multicolumn{1}{c|}{\textbf{Sn}} &
  \multicolumn{1}{c|}{\textbf{Sp}} &
  \multicolumn{1}{c|}{\textbf{Acc}} &
  \multicolumn{1}{c|}{\textbf{AUC}} \\ \cline{2-6} 
 &
  Baseline &
  0.7429 &
  0.9840 &
  0.9544 &
  0.9686 \\ \cline{2-6} 
 &
  Baseline+GLRDN &
  0.7914 &
  0.9791 &
  0.9561 &
  0.9740 \\ \cline{2-6} 
\end{tabular}
\end{table}

\begin{table}[!t]
  \renewcommand{\arraystretch}{1.5}
\caption{Comparison of architectural performance without and with our proposed Euler Characteristic-based Regularizer on the DRIVE datasets.}
\label{tbl-eu}
\begin{tabular}{l|l|l|l|l|l|}
\cline{2-6}
 &
  \multicolumn{1}{c|}{\textbf{Methods}} &
  \multicolumn{1}{c|}{\textbf{Sn}} &
  \multicolumn{1}{c|}{\textbf{Sp}} &
  \multicolumn{1}{c|}{\textbf{Acc}} &
  \multicolumn{1}{c|}{\textbf{AUC}} \\ \cline{2-6} 
 &
  Baseline &
  0.7583 & 
  0.9826 &
  0.9551 & 
  0.9691
   \\ \cline{2-6} 
 &
  Baseline+EC &
  0.8463 &
  0.9759 &
  0.9600 &
  0.9824 \\ \cline{2-6} 
\end{tabular}
\end{table}

\begin{figure*}[!ht]
    \centering
    \includegraphics[width=\textwidth]{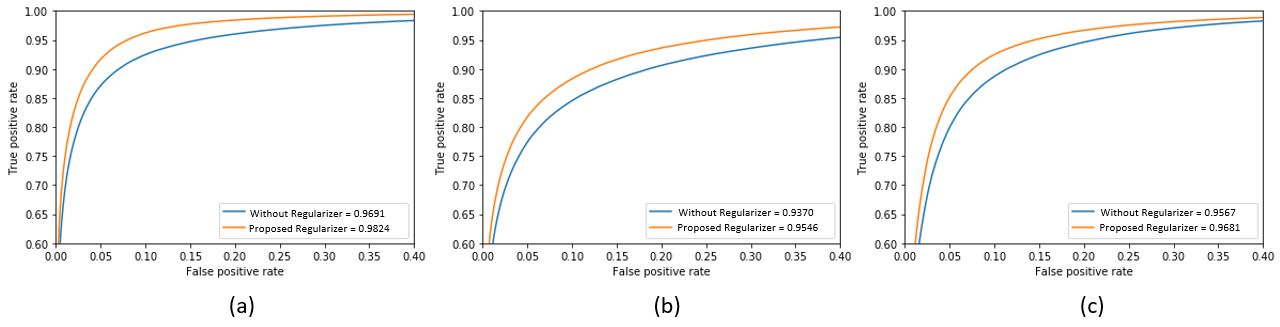}
    \caption{Receiver Operating Characteristics (ROC) curve analysis of blood vessel segmentation between our proposed regularizer and without proposed regularizer on the different datasets. (a)DRIVE (b)STARE and (c)CHASEDB1.}
    \label{fig-roc}
\end{figure*}

\section{Results}
We show some segmentation results from the experiment in the figure \ref{fig-result}. The figure shows a very complicated retinal vessel consisting of large and small vessels crossing each other. The segmentation results from the U-Net baseline appear poor in detecting complex vessels and small vessels. However, the proposed regularizer can improve the segmentation results on small vessels with more detail and better.We also present the results of our quantitatively proposed regularizer on the DRIVE dataset in the \ref{tbl-gbs}, \ref{tbl-glrdn}, and \ref{tbl-eu} tables. Without a regularizer, the baseline network loses a lot of small vessels and creates many isolated objects. These tables indicate that our regularizer is able to segment small vessels and minimize the number of isolated objects on the vessel than without using a regularizer. Figure \ref{fig-roc} shows the performance of our proposed regularizer compared to the baseline network without regularizer on different datasets using ROC curve analysis. In the figure, the blue line shows the ROC curve of the baseline network without the regularizer. The orange line shows the ROC curve of the baseline network merged with our regularizer. The two ROC curve lines show that our proposed regularizer can improve the performance of the baseline network. This indicates that our proposed performance regularizer can provide better performance enhancement on retinal blood vessel segmentation. To get better visualization, we zoomed in on the ROC curve at the top left.

\section{Conclusions}
This study takes advantage of the neighboring pixel information in the image as additional information on the CNN. This information is constructed as a regularization term in the objective function. The method we propose is able to improve vessel segmentation, especially on small vessels, better than without using a regularizer. In addition, isolated objects are also getting minimal. The calculation of regularizer on a pixel-by-pixel basis is a limitation of this approach. In the future, we plan to expand the work by using sparse graphs to reduce computational costs.





\begin{thebibliography}{00}
\bibitem{Ronneberger2015} O. Ronneberger, P. Fischer, and T. Brox, “U-Net: Convolutional Networks for Biomedical Image Segmentation,” in Lecture Notes in Computer Science, Springer International Publishing, 2015, pp. 234–241.

\bibitem{Badrinarayanan2017}V. Badrinarayanan, A. Kendall, and R. Cipolla, “SegNet: A Deep Convolutional Encoder-Decoder Architecture for Image Segmentation,” IEEE Transactions on Pattern Analysis and Machine Intelligence, no. 12, pp. 2481–2495, Dec. 2017, doi: 10.1109/tpami.2016.2644615.

\bibitem{Fu2016}H. Fu, Y. Xu, S. Lin, D. W. Kee Wong, and J. Liu, “DeepVessel: Retinal Vessel Segmentation via Deep Learning and Conditional Random Field,” in Medical Image Computing and Computer-Assisted Intervention – MICCAI 2016, Springer International Publishing, 2016, pp. 132–139.

\bibitem{Havaei2017}M. Havaei et al., “Brain tumor segmentation with Deep Neural Networks,” Medical Image Analysis, pp. 18–31, Jan. 2017, doi: 10.1016/j.media.2016.05.004.

\bibitem{Zeebaree2019}D. Q. Zeebaree, H. Haron, A. M. Abdulazeez, and D. A. Zebari, “Machine learning and Region Growing for Breast Cancer Segmentation,” 2019 International Conference on Advanced Science and Engineering (ICOASE), Apr. 2019, doi: 10.1109/icoase.2019.8723832.


\bibitem{Chougula2020}B. Chougula, A. Tigadi, P. Manage, and S. Kulkarni, “Road segmentation for autonomous vehicle: A review,” 2020 3rd International Conference on Intelligent Sustainable Systems (ICISS), Dec. 2020, doi: 10.1109/iciss49785.2020.9316090.


\bibitem{Ullah2018}M. Ullah, A. Mohammed, and F. Alaya Cheikh, “PedNet: A Spatio-Temporal Deep Convolutional Neural Network for Pedestrian Segmentation,” Journal of Imaging, no. 9, p. 107, Sep. 2018, doi: 10.3390/jimaging4090107.


\bibitem{Cholakkal2019}H. Cholakkal, G. Sun, F. Shahbaz Khan, and L. Shao, “Object Counting and Instance Segmentation With Image-Level Supervision,” 2019 IEEE/CVF Conference on Computer Vision and Pattern Recognition (CVPR), Jun. 2019, doi: 10.1109/cvpr.2019.01268.


\bibitem{Wang2017}C. Wang, Y. Tang, X. Zou, W. SiTu, and W. Feng, “A robust fruit image segmentation algorithm against varying illumination for vision system of fruit harvesting robot,” Optik, pp. 626–631, Feb. 2017, doi: 10.1016/j.ijleo.2016.11.177.

\bibitem{hakim2019}L. Hakim, N. Yudistira, M. Kavitha, and T. Kurita, “U-Net with Graph Based Smoothing Regularizer for Small Vessel Segmentation on Fundus Image,” in Communications in Computer and Information Science, Springer International Publishing, 2019, pp. 515–522.

\bibitem{hakim2021}L. Hakim, M. S. Kavitha, N. Yudistira, and T. Kurita, “Regularizer based on Euler characteristic for retinal blood vessel segmentation,” Pattern Recognition Letters, pp. 83–90, Sep. 2021, doi: 10.1016/j.patrec.2021.05.023.

\bibitem{hakim2022}L. Hakim, H. Zheng, and T. Kurita, “Improvement for Single Image Super-resolution and Image Segmentation by Graph Laplacian Regularizer Based on Differences of Neighboring Pixels,” International Journal of Intelligent Engineering and Systems, no. 1, Feb. 2022, doi: 10.22266/ijies2022.0228.10.




\end{thebibliography}
\end{document}